\documentclass[12pt]{iopart}
\usepackage{ifpdf}
\ifpdf
   \usepackage[pdftex]{graphicx}
   \DeclareGraphicsExtensions{.pdf}
   \usepackage{thumbpdf}      
\else
   \usepackage[dvips]{graphicx}
   \DeclareGraphicsExtensions{.eps}
\fi
\usepackage{cite}
\usepackage{psfrag}
\usepackage{amssymb}
\usepackage{amscd}
\usepackage{subfigure}
\usepackage{setstack}
\usepackage{iopams}
\usepackage{eucal}

\renewcommand{\i}{{\rm i}}

\renewcommand{\text}[1]{{\mbox{#1}}}

\newcommand{\ri}{{ \rm i }}
\newcommand{\re}{{ \rm e }}
\newcommand{\rd}{{ \rm d }}

\newcommand{\be}{\begin{equation}}
\newcommand{\ee}{\end{equation}}

\renewcommand{\vec}[1]{\mathbf{#1}}

\newcommand{\braket}[2]{\langle #1|#2 \rangle}

\newcommand{\kn}[1]{\left( #1 \right)}

\newcommand{\ke}[1]{\left[ #1 \right]}

\renewcommand{\vec}[1]{\mathbf{#1}}
\renewcommand{\i}{\ensuremath{\mathrm{i}}}

\usepackage{color}
\definecolor{blau_s}{rgb}{0,0,1}
\definecolor{blau}{rgb}{0,0,1}
\definecolor{gruen}{rgb}{0,1,0}
\definecolor{rot_s}{rgb}{1,0,0}
\definecolor{rot}{rgb}{1,0,0}
\definecolor{magenta}{rgb}{1,0,1}

\usepackage{enumerate}
\usepackage{exscale}
\usepackage{tabularx}
\usepackage{array}
\usepackage{cite}
\usepackage{listings}

\usepackage{pifont}
\usepackage{stmaryrd}

\begin{document}
\jl{1}
\title[Exceptional points in bichromatic Wannier-Stark systems]
{Exceptional points in bichromatic Wannier-Stark systems}
\author{C Elsen$^1$, K Rapedius$^{1,2}$, D Witthaut$^3$ and H J Korsch$^1$}

\address{$^1$ FB Physik, Technische Universit\"at Kaiserslautern, D-67653
Kaiserslautern, Germany}
\address{$^2$ Center for Nonlinear Phenomena and Complex Systems, Université Libre de Bruxelles (ULB),
Code Postal 231, Campus Plaine, B-1050 Brussels, Belgium}
\address{$^3$ MPI for Dynamics and Self-Organization, Bunsenstra\ss e 10, D-37073 G\"ottingen, Germany}

\eads{\mailto{korsch@physik.uni-kl.de}, \mailto{witthaut@nld.ds.mpg.de }}

\begin{abstract}
The resonance spectrum of a tilted periodic quantum system for a bichromatic
periodic potential is investigated. For such a bichromatic Wannier-Stark system 
exceptional points, degeneracies
of the spectrum, can be localized in parameter space by means of an
efficient method for computing resonances. Berry phases and Petermann factors
are analyzed. Finally the influence of a nonlinearity of the Gross-Pitaevskii type
on the resonance crossing scenario is briefly discussed.
\end{abstract}

\submitto{\JPB}
\pacs{03.65Ge, 03.65Nk, 03.75-b}

\section{Introduction}
\label{s-Intro}
The physics of ultracold atoms and Bose-Einstein condensates in optical 
lattices has made an enormous progress in the last decade. Due to the 
possibility to control all experimental parameters accurately over wide 
ranges and monitor the dynamics of the atoms in situ, optical lattices 
have become one of the most prominent model systems in quantum optics, 
solid state physics and nonlinear dynamics.  Nowadays the shape of the 
optical potential can be engineered with astonishing precision, including 
in particular bichromatic optical lattices \cite{Salg07,Ritt06,Foel07}.  The manipulation 
of matter waves in the lattice is routinely accomplished by a static or 
time-dependent external field, either in an accelerated horizontal 
lattice \cite{Daha96,Ritt06,Salg07,Sias07} or a vertical lattice subject 
to gravity \cite{Ande98,Gust08a,Gust10,Beau11}
also supported by magnetic levitation \cite{Gust08a,Gust10}. 
A weak static field accelerates the atoms up to 
the edge of the Brillouin zone, where they are reflected leading to a 
periodic motion called Bloch oscillation. A strong field 
introduces decay by repeated Landau-Zener tunneling to higher bands. 
Thus optical lattices also became an important model system for the 
study of decay in open quantum systems \cite{Ande98,Ritt06,Sias07,Salg07}.

The dynamics of a quantum particle in a tilted periodic structure has 
been a subject of intensive theoretical investigations starting from 
Bloch's seminal paper on electrons in crystals \cite{Bloc28}. The basic structure 
of the spectrum and the dynamics was then clarified in a discussion by 
Zak and Wannier \cite{Zak68,Zak69,Wann69}: In particular, the spectrum is continuous 
with embedded resonance eigenstates -- the so-called Wannier-Stark 
resonances. These eigenstates are arranged in ladders, where the climbing 
of the ladder is realized by a translation over one lattice period. Each 
ladder can be roughly associated with a Bloch band in the field free case. 
The equidistant spacing of energies in one ladder leads to a fully periodic 
motion --  the celebrated Bloch oscillations. A review of 
these results as well as a surprisingly effective algorithm to calculate 
Wannier-Stark resonance states is given in \cite{02wsrep}.

In the present paper we investigate a rather peculiar feature of tilted 
bichromatic lattices, the existence and properties of exceptional points. 
In an open system the eigenenergies become complex valued, where the imaginary part 
gives the decay rate (for a comprehensive discussion of such non-hermitian
quantum systems see the recent textbook by Moiseyev \cite{Mois11}).
If a system parameter is varied, as for example the external field
strength, the eigenvalues show avoided crossings. For resonance states 
two types of level crossings scenarios exist -- either the 
real parts of the energies anti cross and the imaginary parts cross, denoted
as a type I crossing, or vice versa, a type II crossing \cite{03crossing}.
A full 
coincidence of the real and imaginary parts is possible at isolated 
points in parameter space, the so-called exceptional points (EPs). The existence 
of these points has a remarkable implication on the dynamics of the system. 
Suppose the system parameters are varied adiabatically along a cyclic path 
around an exceptional point. In general, a cyclic evolution leaves the 
quantum state invariant up to a geometric phase, or Berry phase, which is of great importance 
both from a fundamental viewpoint as well as for applications \cite{Shap89,Sjoq08}. 
Cycling around an exceptional point has an even stronger effect: In addition 
to a geometric phase it exchanges the two crossing states.

Both crossings of type I and II
appear for Wannier-Stark systems where the periodic potential is sinusoidal
(see, e.g., \cite{02wsrep}), but exceptional points have not been detected.
A modulated periodic potential with
two additional parameters is, however, flexible enough to allow for such
fascinating degeneracies and first results have been reported  recently  
for a bichromatic potential \cite{10nlret}. Such systems may be
very well suited for experimental studies of the phenomena generated by such
degeneracies for quantum systems, where experiments are still rare. 

In the following we will discuss the energy spectrum of a quantum 
particle in a tilted bichromatic optical lattice in dependence
of the system parameters.
A pronounced feature of such a Wannier-Stark system is the
enhancement of the decay rate by resonant tunneling (RET) when the 
energies of two different Wannier-Stark ladders cross. Most interestingly, 
this crossing can be of both types, with an anti-crossing of either the 
energy or the decay rate, and even give rise to an exceptional point. The 
properties of these points and their embedding in the energy 
surfaces is discussed in detail. Exceptional points are most clearly 
identified by the divergence of the Petermann factor, which measures the 
self-overlap of right and left eigenvectors of the same state. Thus one 
can actually use this quantity for a systematic search for exceptional 
points in parameter space. 

Finally we extend our studies to the stationary 
states of the nonlinear Schr\"odinger equation, which describes the dynamics 
of a Bose-Einstein condensate in a mean-field approach. The nonlinearity has 
a dramatic effect on the RET peaks of the decay rate, bending their shape 
and shifting their position \cite{08nlLorentz}. Thus it obviously strongly influences the 
exceptional points, too.
 
\section{Wannier-Stark resonances}
\label{sec-WS}
As mentioned in the introduction, we will demonstrate the existence of exceptional
points for a tilted periodic potential with modulated potential minima and maxima, a
bichromatic potential. However,
before we turn to the bichromatic case we first review some general properties of the Wannier-Stark Hamiltonian
\be
   H=-\frac{\hbar^2}{2m} \frac{\rd^2}{\rd x^2}+V(x)+Fx\,,
\label{H_WS}
\ee
where the potential $V(x+d)=V(x)$ has a single period $d$.
It was shown \cite{Zak68,Wann69,Zak69,Avro77} that the Hamiltonian (\ref{H_WS})
with $F > 0$ has a continuous spectrum in which discrete ladders
\be
   {\mathcal E}_{\alpha,n}=E_{\alpha,0}+n d F -\ri \Gamma_\alpha/2
\label{WS_ladder}
\ee
of complex valued resonances are embedded
, where $\alpha=1,2, ...$ is the ladder index and $n$ is the site index. The decay rates 
$\Gamma_\alpha$, which are the same for all resonances
in one ladder, are due to the finite probability for tunneling out of the lattice toward 
$x \rightarrow -\infty$ where the Stark term $Fx$ goes to $-\infty$ (for positive fields 
$F>0$) so that there is no reflection.
The resonance energies (\ref{WS_ladder}) and corresponding wavefunctions thus satisfy the nonhermitean 
eigenvalue problem 
\be 
   H \Psi_{\alpha,n}(x)= {\cal E}_{\alpha,n}\Psi_{\alpha,n}(x)
\ee
with purely outgoing (Siegert) boundary conditions 
$\partial_x \Psi_{\alpha,n} (x)\rightarrow -\ri k(x) \Psi_{\alpha,n}(x)$ 
for $x \rightarrow -\infty$ with the local wavenumber
$k(x)=\sqrt{2m({\cal E}_{\alpha,n}-V(x)-Fx)}/\hbar$. At $x \rightarrow \infty$ the wavefunction 
satisfies the usual bound state condition $|\Psi_{\alpha,n}(x)|^2 \rightarrow 0$.
Equivalent ways of avoiding reflection at $x \rightarrow -\infty$ include the use of complex 
absorbing potentials (see \cite{Mois11,10nlret} and section \ref{sec-nlcross}) or complex scaling 
of the coordinate 
$x$ \cite{Mois11,98wannier}. Here, we use an efficient calculation method introduced in 
\cite{99trunc,98wannier} based on a finite basis expansion in momentum space.

The Wannier-Stark ladder (\ref{WS_ladder}) can be understood by considering the commutator
\be
    [H, T_\ell]=-\ell d F T_\ell
     \label{H_WS_Komm}
\ee
between the Wannier-Stark Hamiltonian (\ref{H_WS}) and the translation operator $T_\ell$ over $\ell$ lattice sites.
Equation (\ref{H_WS_Komm}) expresses the fact that translation by $\ell$ lattice sites has the same effect as an 
energy shift $\ell d F$. Thus we obtain
\be
\fl   \qquad H T_\ell \Psi_{\alpha,n}(x)= T_\ell H\Psi_{\alpha,n}(x)+[H,
T_\ell]\Psi_{\alpha,n}(x)
    =( {\cal E}_{\alpha,n}- \ell d F) T_\ell \Psi_{\alpha,n}(x)\,,
\ee
which leads to the ladder (\ref{WS_ladder}) of resonance energies with the respective wavefunctions
\be
   \Psi_{\alpha,n}(x)=\Psi_{\alpha,0}(x-n d F)\,. 
\ee

The general dependence of the decay rates on the field strength is approximately given by 
$\Gamma (F)\propto F \exp(-\pi \Delta E^2/F)$ where $\Delta E$ is the energy gap between the
ground and first excited Bloch band. This result is obtained by Landau--Zener theory under the assumption that decay 
is mainly determined by tunneling from the ground band to the first excited band as successive tunneling 
events into higher bands, which finally leads to decay towards $x \rightarrow -\infty$ are fast compared 
to the first tunneling process \cite{Holt00,00restun}. Deviations from the Landau-Zener dependence occur 
if a state of a lower ladder with energy ${\cal E}_{\alpha,n}$ is in resonance with  a state of a higher 
ladder at a different site, i.~e.~$E_{\alpha,n}=E_{\alpha',n'}$ 
\cite{Sias07,02wsrep}. Such a resonant coupling between two 
ladders leads to a strong increase of the decay rate of the lower ladder so 
that it is called resonantly enhanced tunneling (RET). At the same time, the decay rate in the upper ladder 
is lowered.
Whenever two ladders are in resonance, the decay rate for the lower ladder shows a peak whereas the upper band has a dip.

Now we turn to the bichromatic or double--periodic potential
\begin{equation}
V(x) = \frac{V_0}{2} \kn{ \cos(2\pi x/d) + \delta \cos(4\pi x/d+\phi) },
\label{math-mws1-pot}
\end{equation}
given by the sum of a grid with period $d$ and an additional $d/2$-periodic grid which
creates $d$-periodic potentials with modulated minima and maxima of different
heights, depending on $\phi$.
Throughout this paper we use scaled units such that $d$ is equal to $2\pi$ and $m=\hbar=1$.
The energies are measured in units of $8E_R$, where $E_R=\hbar^2\pi^2/(2md^2)$ is the
recoil energy. In the subsequent computations, we furthermore fix the potential strength as $V_0=1$.

The band structure of the double-periodic potential (\ref{math-mws1-pot}) 
is characterized by the splitting of the ground band into two 
\textit{minibands}. The 
Wannier-Stark ladder splits into two \textit{miniladders}, where the  width of 
the splitting depends on the modulation $\delta$. In the limit of vanishing 
modulation $\delta \rightarrow 0$ the miniladders are energetically degenerate. 
In analogy to equation (\ref{H_WS_Komm}) one can derive the relations
\begin{eqnarray}
 \ke{H,T_{2\ell}} &= -2\ell dFT_\ell \label{math-mws1-kom1}\\
 \ke{H,T_{2\ell+1}G} &= -(2\ell+1)dFT_\ell
\label{math-mws1-kom2}
\end{eqnarray}
(see \cite{06bloch_zener,06bloch_manip} for details),
where $H$ is the Hamiltonian, $T_\ell$ the translation operator over $\ell$ 
lattice sites and $G$ is an operator that switches the sign of the modulation 
$\delta$  in all following terms.
 Using (\ref{math-mws1-kom1}) 
and (\ref{math-mws1-kom2}) one can show that the Wannier-Stark ladder 
of eigenenergies of the unperturbed system splits into the two 
miniladders: 
\begin{eqnarray}
 \mathcal{E}_{\alpha,2\ell} &= \hphantom{-}\mathcal{E}_{\alpha}(\delta) + 2\ell dF\\
 \mathcal{E}_{\alpha,2\ell+1} &= -\mathcal{E}_{\alpha}(\delta) + (2\ell+1)dF.
\end{eqnarray}
The energy offset $\mathcal{E}_{\alpha}(\delta)$ is an antisymmetric function of $\delta$.
Each Wannier-Stark ladder bifurcates into two which are energetically shifted
with a distance $2\mathcal{E}_{\alpha}(\delta)$.

\begin{figure}[bth]
\centering
 \includegraphics[height=8cm, angle=0]{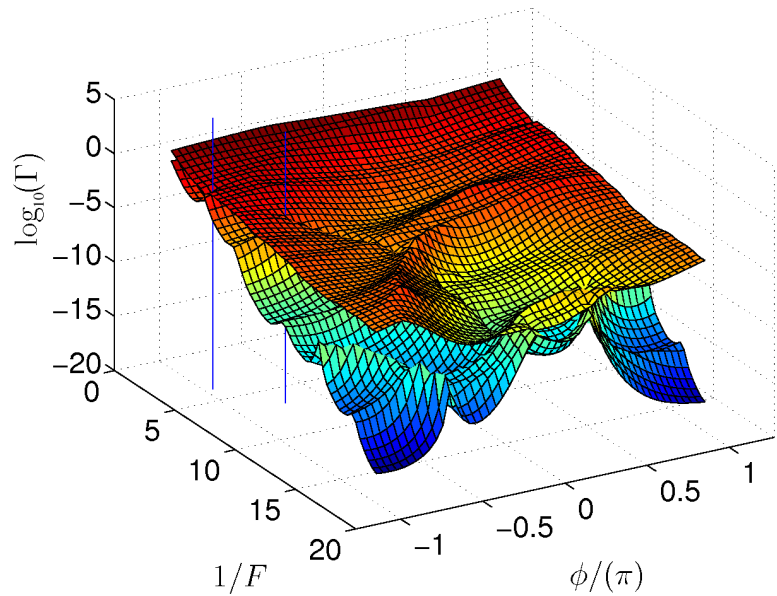}
 \caption{Decay rate $\Gamma_{1,2}$ as a function of the inverse field strength $1/F$ and the phase $\phi$ with
          $\delta=1$. The configurations marked by a vertical line correspond to EPs.}
 \label{fig-mws1-riemann_d1_imag}
\end{figure}

Let us first consider the topology of the energy surfaces
in an arbitrarily chosen part of the parameter space. 
Figure \ref{fig-mws1-riemann_d1_imag} shows the imaginary part of the eigenenergies, i.e.~the decay rate
 of the two most stable states in dependence on the two parameters $\phi$ and $1/F$ 
 where the modulation $\delta=1$ is kept fixed.

If we vary only $1/F$ keeping the potential, i.e. $\phi$, fixed, we observe the familiar RET spectra 
with eigenvalue crossings and avoided crossings.
For the cut with  $\phi=-1$ appearing at the front of the surfaces in figure \ref{fig-mws1-riemann_d1_imag}
for example, we have a series of avoided crossings of the imaginary parts, whereas the real parts cross,
i.e.~a type II crossing scenario. Cuts at different values of $\phi$ show type I crossings. One is tempted
to anticipate to find fully degenerate eigenvalues somewhere in the $(1/F,\phi)$-plane. This will be clarified
in the following section.

\section{Exceptional points}
\label{sec-EP}
\subsection{Overview}
Exceptional points occur in systems described by non-hermitian Hamiltonians 
depending on a set of parameters.
As mentioned above, a point in parameter 
space at which both the complex eigenvalues and the eigenstates of two 
different eigenmodes coincide is called an  exceptional point
\cite{Kato66book,Heis00,Mois11}. EPs must be distinguished from so-called 
diabolic points  that occur if there is a coincidence 
of the eigenvalues but not of the eigenstates.  Diabolic points are more familiar from hermitian systems, but in fact
they are more special than the exceptional points and can be viewed
as a coincidence of two exceptional points (see, e.g., \cite{03crossing,Berr03} and references
therein). 
Bichromatic optical lattices showing
diabolic points in the band structure have been proposed as quantum 
simulators for the Dirac equation \cite{Witt11}.

\begin{figure}[htb]
\centering
\includegraphics[height=6.0cm, angle=0]{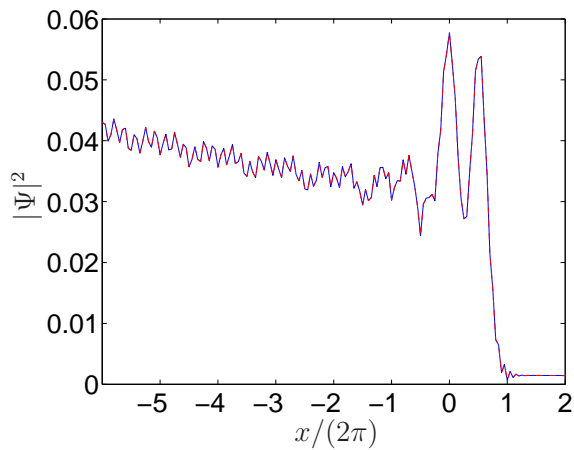}
\caption{Wannier-Stark eigenfunctions $\textcolor{blau}{\Psi_1}$ and $\textcolor{rot}{\Psi_2}$ in position space
         for the configuration of the EP (\ref{math-mws1-ep}). The wavefunctions coincide on the scale of drawing.}
\label{fig-mws1-d1_x}
\end{figure}

Before we introduce a systematic algorithm for finding
the exceptional points for the tilted periodic potential (\ref{math-mws1-pot}), 
we first have a look at
the set of solutions $\mathcal{E}(1/F,\delta,\phi)$ shown in figure \ref{fig-mws1-riemann_d1_imag}.
Exceptional points are searched by means of the criterion that
the energy difference $\Delta \mathcal{E} = |\mathcal{E}_{1}-\mathcal{E}_{2}|$ between the two lowest levels must vanish.
This criterion is satisfied for the following points marked by vertical lines in the figure: 
\begin{eqnarray}
{\; 1/F=3.769,\quad\delta=1,\quad\phi=-2.991, \;}
\label{math-mws1-ep}\\
{\; 1/F=6.662,\quad\delta=1,\quad \phi=-2.228. \;}
\label{math-mws1-ep-1}
\end{eqnarray}
One can straightforwardly verify that the two eigenfunctions involved are also degenerate for these 
parameters so that the points (\ref{math-mws1-ep}) and (\ref{math-mws1-ep-1}) are indeed 
exceptional points. This is illustrated in figure \ref{fig-mws1-d1_x} which exemplarily shows 
the two Wannier-Stark eigenfunctions $\Psi_1$ and $\Psi_2$ for the parameters (\ref{math-mws1-ep}). One observes 
that the wavefunctions are localized in two neighboring potential minima at $x=0$ and $x=0.5$.

\subsection{Emergence of exceptional points in tilted optical lattices}
In the following we want to address the question how the emergence of 
exceptional points in a bichromatic Wannier-Stark potential can be 
understood in terms of the shape of the potential. The definition of an 
EP requires the coincidence of both, the energies and eigenfunctions of 
the two states considered.

Let us first recall the features of the single-periodic Wannier-Stark 
system introduced in section \ref{sec-WS}, where a simultaneous degeneracy of the 
eigenenergies and eigenfunctions is impossible. The Stark potential $Fx$ 
leads to a localization of the eigenfunctions and induces a ladder 
structure of the eigenenergies $E_{\alpha,n} = E_{\alpha,0} + 2\pi F n$ 
(see eq.~(\ref{WS_ladder})). For a finite field $F > 0$ the energies within a ladder 
differ by multiples of $2\pi F$, such that a degeneracy in $E$ can only 
occur between states of different ladders, resulting in resonantly 
enhanced tunneling. However, the states involved are localized in 
different wells of the potential, such that they cannot be described by 
the same wavefunction.

\begin{figure}[b]
\centering
\includegraphics[height=5.7cm, angle=0]{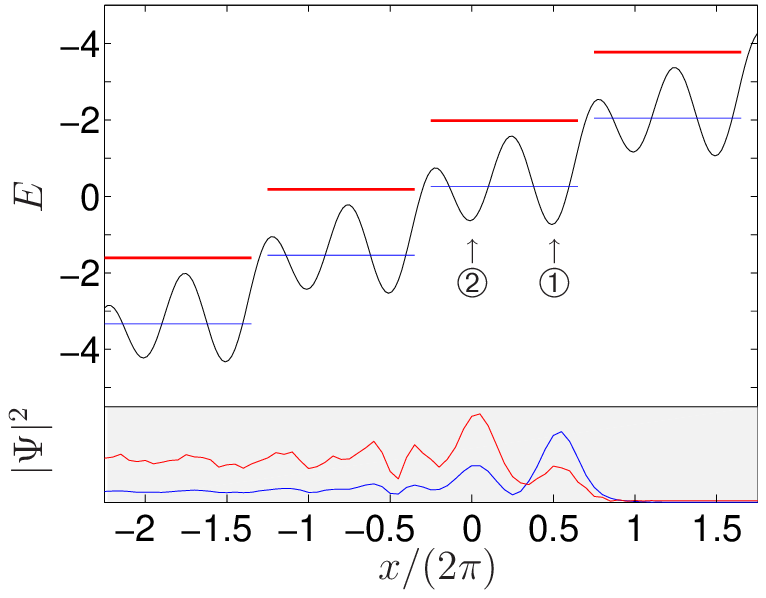}
\includegraphics[height=5.7cm, angle=0]{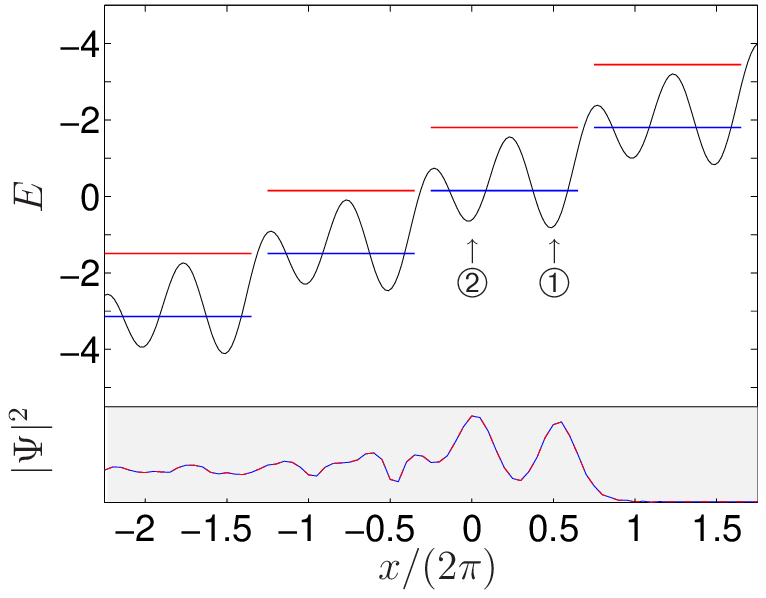}
\caption{Configuration in the vicinity of the EP 
(left) and exactly at the EP (\ref{math-mws1-ep2}) (right). The
upper panels show the Wannier-Stark ladders for the two most stable resonances, where
the widths of the lines indicate the decay probability. The lower panels show the
Wannier-Stark functions \textcolor{blau}{$\Psi_1$ (blue)} and \textcolor{rot}{$\Psi_2$ (red)}.} 
\label{fig-mws1-dpexample}
\end{figure}

The situation is different in a bichromatic lattice. Within one period 
of the potential $V(x)$, there are two potential minima, in which a 
Wannier-Stark resonance is localized. By varying one parameter, e.g.~the 
field strength $F$, one can achieve a degeneracy of the eigenenergies. 
Now it is possible to fine-tune also the remaining parameters to realize 
a full coincidence as illustrated in figure \ref{fig-mws1-dpexample}.
An exceptional point is 
found when the more stable state is destabilized and vice versa while 
keeping their energies degenerate.

\begin{figure}[t]
\centering
\includegraphics[height=5.5cm, angle=0]{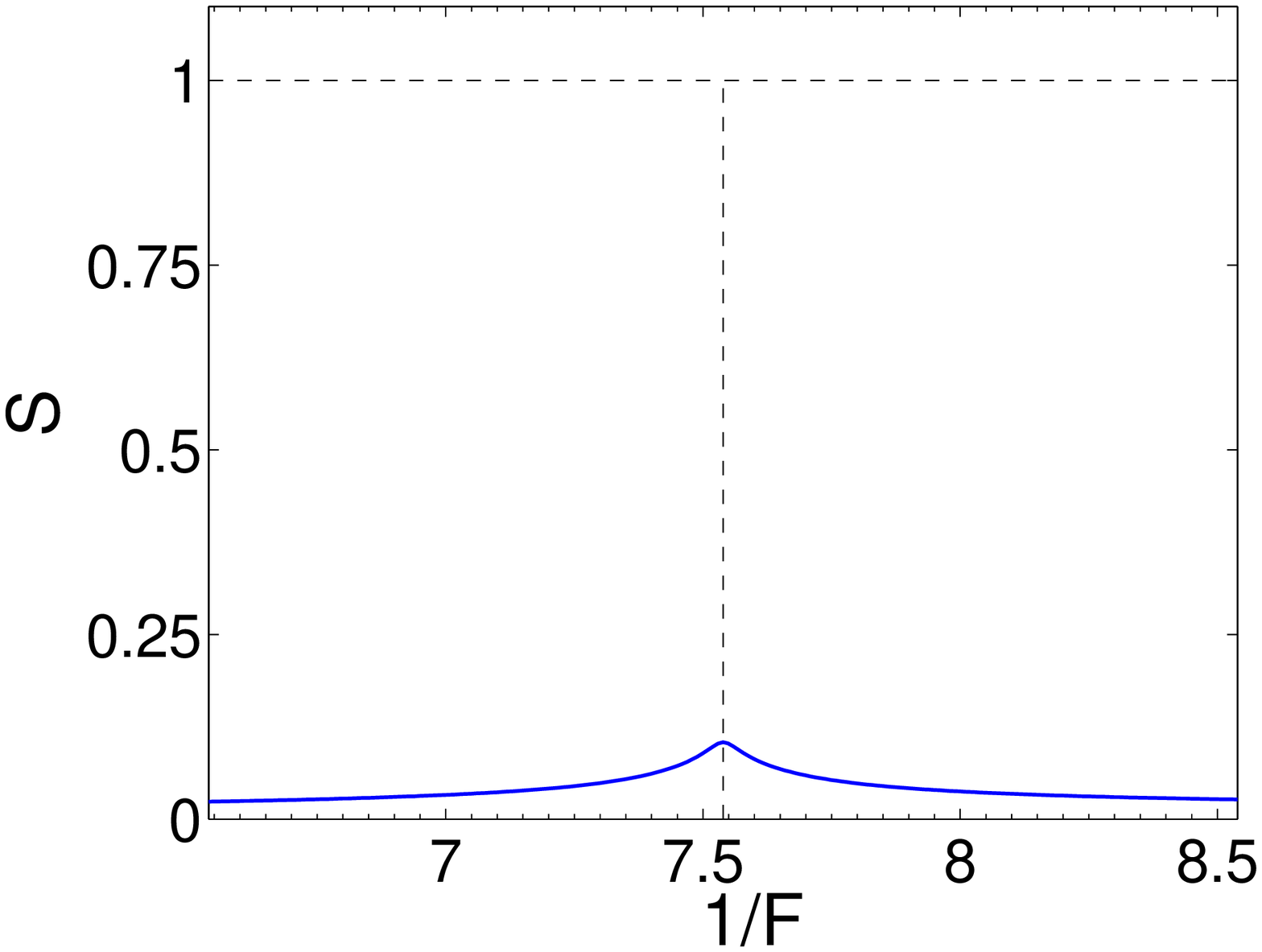}
\hspace*{2mm}
\includegraphics[height=5.5cm, angle=0]{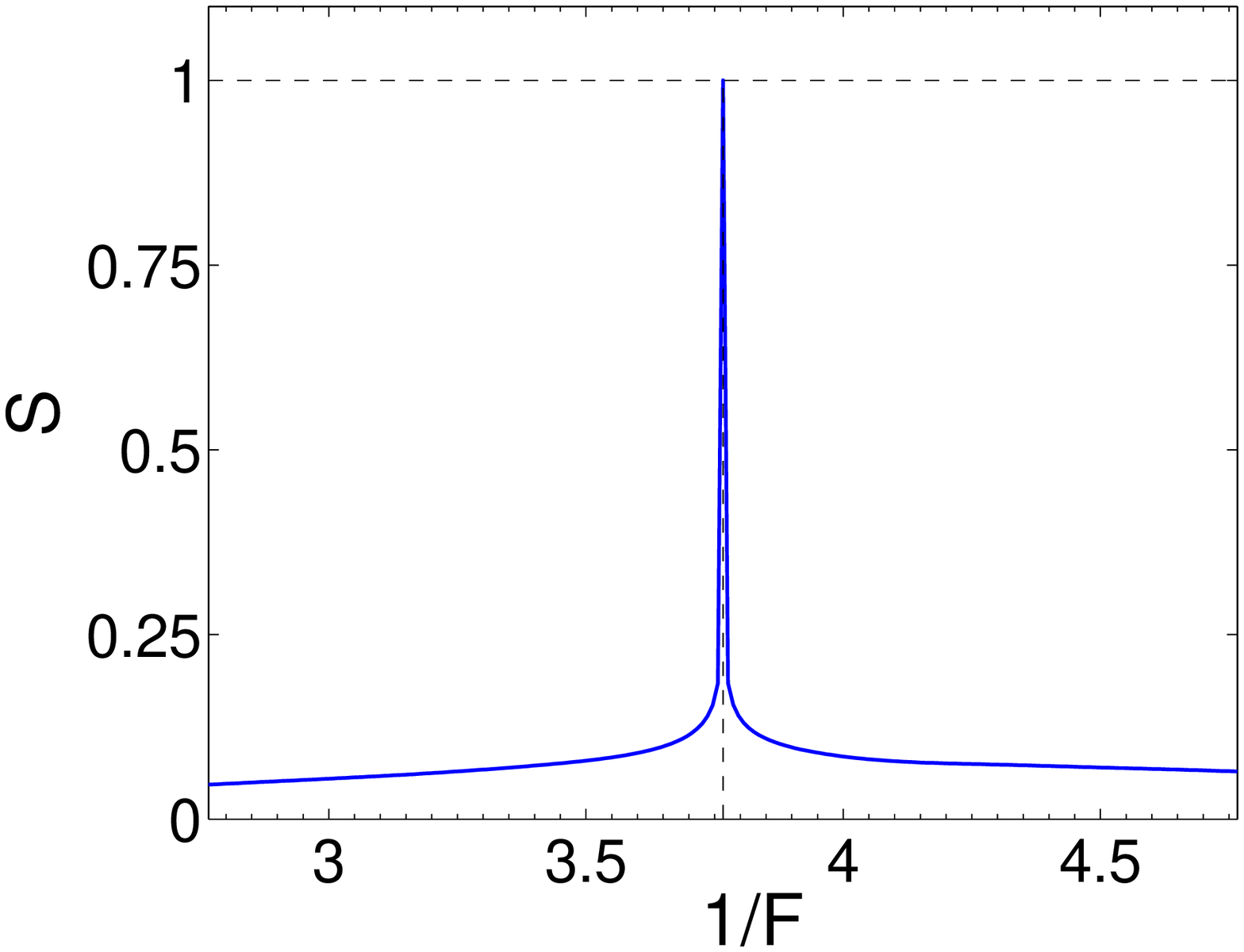}
\caption{Value of the overlap integral $S$ for a variation of the field strength $F$ in the case of
a single-periodic (left panel) and a double-periodic potential (right panel). The
dashed horizontal line indicates the maximum value $S=1$, the vertical line indicates the position of the maximum
at $1/F=7.5$  respectively $1/F=3.8$.} 
\label{fig-mws1-overlap}
\end{figure}

At the exceptional point, both the eigenenergies and the eigenstates 
must coincide. Therefore the overlap integral
\be
S_{1,2}=\big|\braket{\Psi_1}{\Psi_2}\big|\le 1
\label{math-mws1-overlapI}
\ee
of the two (normalized) eigenstates should approach unity at the EP. 
For parameters far away from it the overlap is expected to be small. 
As an example, figure \ref{fig-mws1-overlap} displays the overlap 
in dependence on $1/F$ for system parameters at some distance from an exceptional point (left
panel), where we find a rather small peak of the overlap.
The right panel of figure \ref{fig-mws1-overlap} displays the overlap $S$ in 
dependence on $F$ in the vicinity of the EP  (\ref{math-mws1-ep}). For $1/F \approx 3.7$, which corresponds 
to the configuration of the EP, we observe a pronounced peak with  $S \approx 1$ as the two wavefunctions 
are degenerate. This behavior can be used as a tool for detecting an EP.
Note that
alternatively one might consider the Petermann factor $K_{\alpha}$ describing the 
`(self)overlap' between left and right eigenvectors that we will use in section \ref{sec-nlcross}.

To further investigate this phenomenon, figure \ref{fig-mws1-dpexample} shows both the potential and the wavefunction
for the parameters
\begin{equation}
 {\; 1/F=3.814,\quad\delta=2.251,\quad \phi=-3.035. \;}
\label{math-mws1-ep2}
\end{equation}
that correspond to an EP as well as for a slightly different parameter set in the vicinity of the EP (see table 
\ref{tab-mws1-ep2} for the exact configurations). In agreement with our previous observations, the overlap only 
assumes values of $S \approx 1$ for a very narrow interval of field strengths $F$. The potentials shown in the 
two panels in figure \ref{fig-mws1-dpexample} are hardly distinguishable at first glance.
Yet there is a significant difference as far as the corresponding pairs of Wannier-Stark eigenfunctions and 
in particular their overlap are concerned. The difference is also evident in the corresponding complex eigenenergies 
given in table \ref{tab-mws1-ep2}. 
\begin{table}[hbt]
\caption{\label{tab-mws1-ep2} System parameters sets in the vicinity of the EP
(\ref{math-mws1-ep2}).}
\begin{indented}
\item[]\begin{tabular}{@{}ccc|cccc}
\br
 $1/F$ & $\delta$ & $\phi$ & $E_1$ & $\Gamma_1$ & $E_2$ & $\Gamma_2$\\
 \hline
 3.000 & 2.251 & -3.141 & 2.561 $\times 10^{-1}$ & 2.961 $\times 10^{-2}$ & 1.858 $\times 10^{-1}$ &
1.361 $\times 10^{-1}$\\
 3.814 & 2.251 & -3.035 & 1.538 $\times 10^{-1}$ & 7.427 $\times 10^{-2}$ & 1.538 $\times 10^{-1}$ &
 7.427 $\times 10^{-2}$\\
 \br
\end{tabular}
\end{indented}
\end{table}
Looking at the decay rates we observe that at the transition to the EP the ground state becomes less stable whereas the lifetime of the excited state increases. 

The emergence of the degeneracy can be related to the shape of the double-periodic potential.  Within a 
period of the potential $V(x)$ we have two minima marked by \ding{192} and \ding{193}
in the left lower panel of figure \ref{fig-mws1-dpexample}. 
These minima are slightly shifted with respect to each other. For
$V_{\mathrm{min}}^{\text{\ding{192}}} < V_{\mathrm{min}}^{\text{\ding{193}}}$ the ground state wavefunction is mainly 
localized in the first well  (cf.~left panel of figure \ref{fig-mws1-dpexample}). Together 
with well \ding{193} one obtains the shape of a double well potential. The first excited state reveals 
a high probability density in the second well. Since the barrier in the direction $x \rightarrow -\infty$ 
is lower for well \ding{193}, the probability for tunneling processes in this direction is higher 
leading to a larger decay rate (cf.~table \ref{tab-mws1-ep2}).

For the field strength $1/F \approx 3.814$ corresponding to the configuration of
the EP, the ground and first excited resonance state are energetically  fully
degenerate. The first excited state is stabilized since tunneling towards the ``more stable well'' \ding{192} is enhanced.
At the same time the ground state is destabilized because the probability for tunneling from well \ding{192} toward $x \rightarrow -\infty$ is likewise enhanced. The EP corresponds to the balanced situation where the two Wannier-Stark states become indistinguishable so that there is a degeneracy between wavefunctions belonging to different bands which cannot be achieved in the single-periodic Wannier-Stark system.

\subsection{Systematic search for exceptional points}
\label{subsec-systematic}
So far all EPs have been determined in the same way: If one suspects the existence of an EP in some part of the 
configuration space, eigenvalues and 
eigenfunctions of the Hamiltonian are analyzed with respect to the conditions
\begin{eqnarray}
 && |\mathcal{E}_1-\mathcal{E}_2|=0 \label{math-mws1-krit1} \\
 && S_{1,2}=\big|\braket{\Psi_1}{\Psi_2}\big|= 1\,. \label{math-mws1-krit2}
\end{eqnarray}
In the present study we numerically minimized the eigenvalues distance in (\ref{math-mws1-krit1})
 using a standard method based on the  \emph{simplex algorithm} 
 (see, e.~g.~\cite{FundOpt,lagarias:112}). Finally it was 
 checked if the overlap condition (\ref{math-mws1-krit2}) is satisfied as well. 
 Note that a more efficient methods for localizing exceptional
points have been developed recently \cite{Lefe10,Uzdi10}. 

So far, one of the three parameters $(1/F,\delta,\phi)$ has been kept fixed whereas 
the other two were varied.  
Now we want to search for EPs in the fully three-dimensional parameter space by 
solving the minimization problem (\ref{math-mws1-krit1}). 
In order to converge to the desired EP instead of some local minimum the algorithm requires a suitable initial 
guess for the parameters
$(1/F,\delta,\phi)$. To this end we represent the parameter space in terms of spherical coordinates 
\begin{equation}
\eqalign{
    \Delta (1/F) &= r \sin \vartheta \cos \varphi \\
    \Delta \delta &= r \sin \vartheta \sin \varphi \\
    \Delta \phi &= r \cos \vartheta.
    }
\label{math-mws1-coordtrans}
\end{equation}

If the parameters 
$\vec{r}_{\mathrm{EP}}=(1/F_{\mathrm{EP}},\delta_{\mathrm{EP}},\phi_{\mathrm{EP}})$ of an EP are known, one can 
look for further solutions $\vec{r}_{\mathrm{EP}}' = \vec{r}_{\mathrm{EP}} +
\Delta \vec{r}$ in its vicinity. Thus we search for EPs in the following manner:
\begin{enumerate}
 \item Find an EP (e.~g.~with the methods described further above);
 \item Choose some small ``distance'' $r$, in which to look for the next EP (see below);
 \item Minimize the condition (\ref{math-mws1-krit1}) using the simplex algorithm with the
   parametrization $(r=\mathrm{const},\vartheta,\varphi)$.
 \item Repeat this procedure starting with the configuration of the newly found EP. 
\end{enumerate}
For the sake of completeness we note that we use 
\begin{equation}
\eqalign{
 r(\Delta(1/F),\Delta \delta, \Delta \phi) =
\sqrt{(\Delta(1/F))^2+(\Delta \delta)^2+(\Delta \phi)^2}
}
\end{equation}
with 
$\Delta(1/F) = 1/F_1-1/F_2$, $\Delta \delta = \delta_1 - \delta_2$ and  $\Delta \phi = \phi_1 - \phi_2$
as the distance between two configurations $(1/F_1,\delta_1,\phi_1)$ and $(1/F_2,\delta_2,\phi_2)$.

With this procedure the minimization problem (\ref{math-mws1-krit1}) is solved in a two-dimensional 
subspace (the surface of a sphere) of the three dimensional parameter space which reduces the numerical effort. 
The finite area of that subspace also improves the convergence.
This systematic search detects all exceptional points between the two 
ladders in the parameter region under consideration. Obviously, further 
exceptional points can exist between higher ladders, which are not 
detected. However, these are of less practical interest, as the decay is 
significantly stronger.

\begin{figure}[htb]
\centering
\begin{minipage}[b]{9cm}
\includegraphics[height=7cm, angle=0]{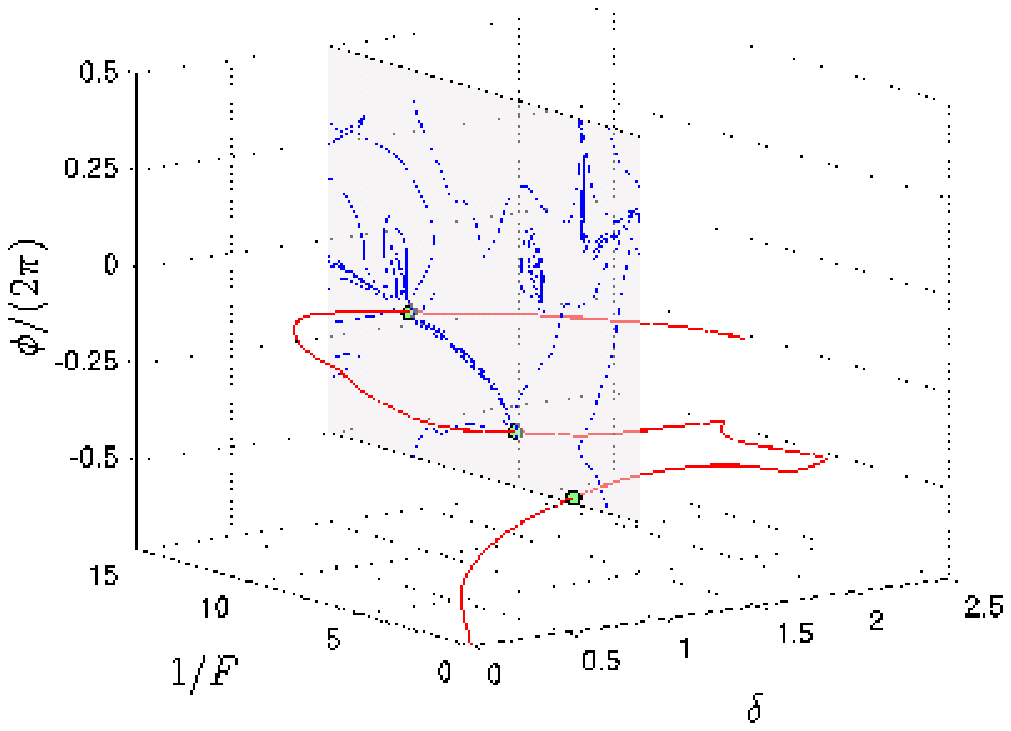}
\end{minipage}
\begin{minipage}[b]{6cm}
\includegraphics[height=5cm, angle=0]{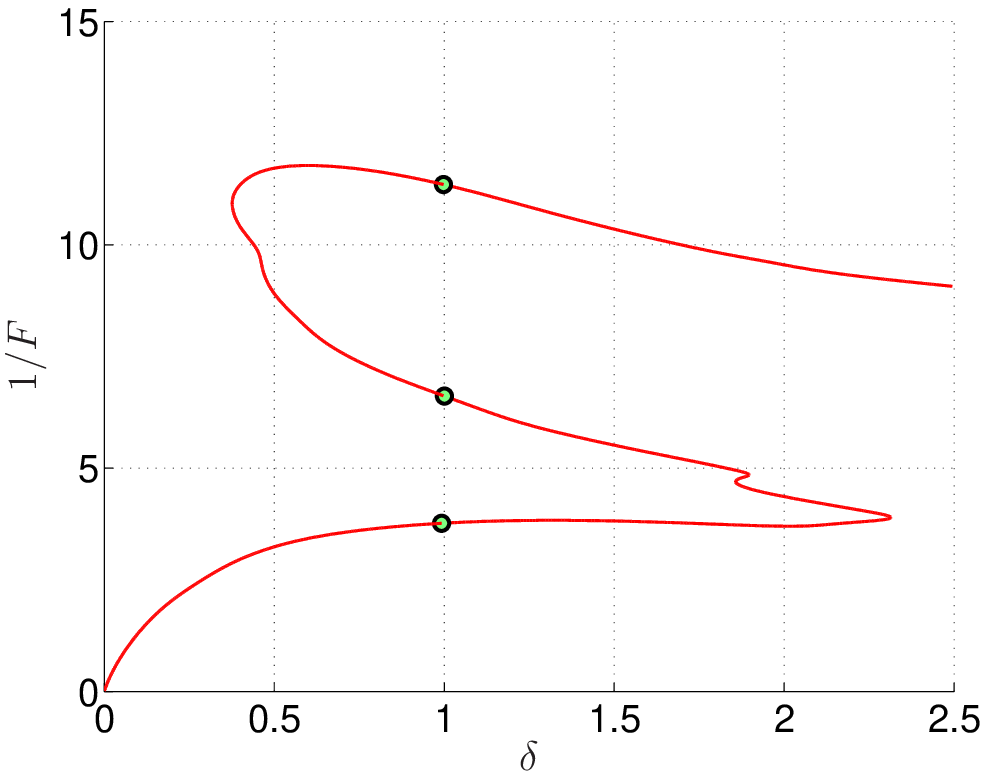}
\vspace*{1mm}
\end{minipage}
\caption{Left panel: Location of the EPs in parameter space for the two most 
stable resonances (red curve). On the plane for $\delta=1$  the absolute value of the difference of the decay 
rates $\Delta \Gamma=|\Gamma_1 - \Gamma_2|$ is shown as a contour plot. Right panel: Curve traced
out by the EPs projected onto the ($1/F$, $\delta$)--plane.}
\label{fig-mws1-energyls}
\end{figure}

Starting with the EP (\ref{math-mws1-ep}) we obtain a one-dimensional manifold formed by the EPs 
of the two most stable resonances of the double-periodic Wannier-Stark system that is embedded 
in the three-dimensional parameter space. This is displayed in figure
\ref{fig-mws1-energyls}.
In addition, we have included a contour plot of the absolute difference $\Delta \Gamma=|\Gamma_1 - \Gamma_2|$ 
of the decay rates on the plane $\delta=1$. The intersections of the EP curve with this plane are located 
at the zeros of (\ref{math-mws1-krit1}). For $\delta=1$, three solutions can be detected
in the region considered.

One can draw the following conclusions:
In the single-periodic system no exceptional points were found. The results for the 
bichromatic system reveal the transition between the two different potentials: In the limit $\delta \rightarrow 0$ the position of the EP is shifted toward an infinitely strong field strength $F$ ($1/F \rightarrow 0$), so that an EP cannot exist in the single-periodic limit. 

Finally we briefly want to discuss the shape of the curve. From figure \ref{fig-mws1-energyls}
we see that the EP curve bends strongly for $\delta \approx 2.3$, so that no EPs can be found for higher values of the modulation $\delta$.
Yet for small field strengths $F$, i.~e.~for high values of $1/F$, there are intersections of the EP curve with the planes
$(1/F,\delta=\mathrm{const},\phi)$. This local behavior repeats itself and can also be observed
for higher bands. 
The results in the domain of interest support our expectation that
there are no EPs in the limit $F \rightarrow 0$, i.e. without an external field.
To gain further insight into the characteristic shape of the EP curve additional investigations are needed.

\subsection{Cyclic parameter variation}
\label{mws1-evol}

In the following we want to study the behavior of eigenvalues and eigenvectors under cyclic parameter variation.
As an example we consider the vicinity of the EP (\ref{math-mws1-ep}) and choose
$\delta=1$ to be constant. The quantities $F$ and $\phi$ are varied along the paths
\begin{equation}
\eqalign{
 1/F  &= 1/F_{\mathrm{EP}}   + r\sin(\beta) + 1/F_0\\
 \phi &= \phi_{\mathrm{EP}}  + r\cos(\beta) + \phi_0
}
\end{equation}
parametrized by an angle $\beta  \in [0,2\pi)$. As in subsection \ref{subsec-systematic} $r$ is a numerical quantity that can be identified as a radius in parameter space.

\begin{figure}[t]
 \centering
 \includegraphics[height=5cm, angle=0]{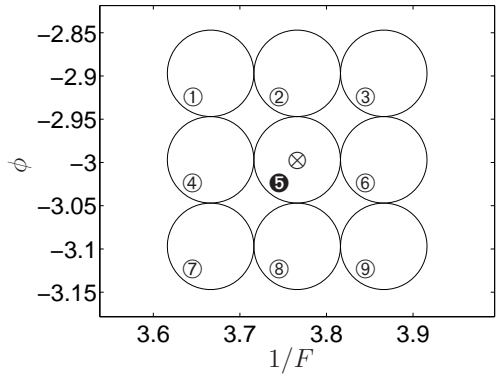}
 \caption{Cyclic variation of the parameters $F$ and $\phi$ for $\delta=1$. The path \ding{206}
          encloses an EP ($\varotimes$), the other paths do not.}
 \label{fig-mws1-EPcircle1-path}
\end{figure}

\begin{figure}[b]
 \centering
 \includegraphics[height=10cm, angle=0]{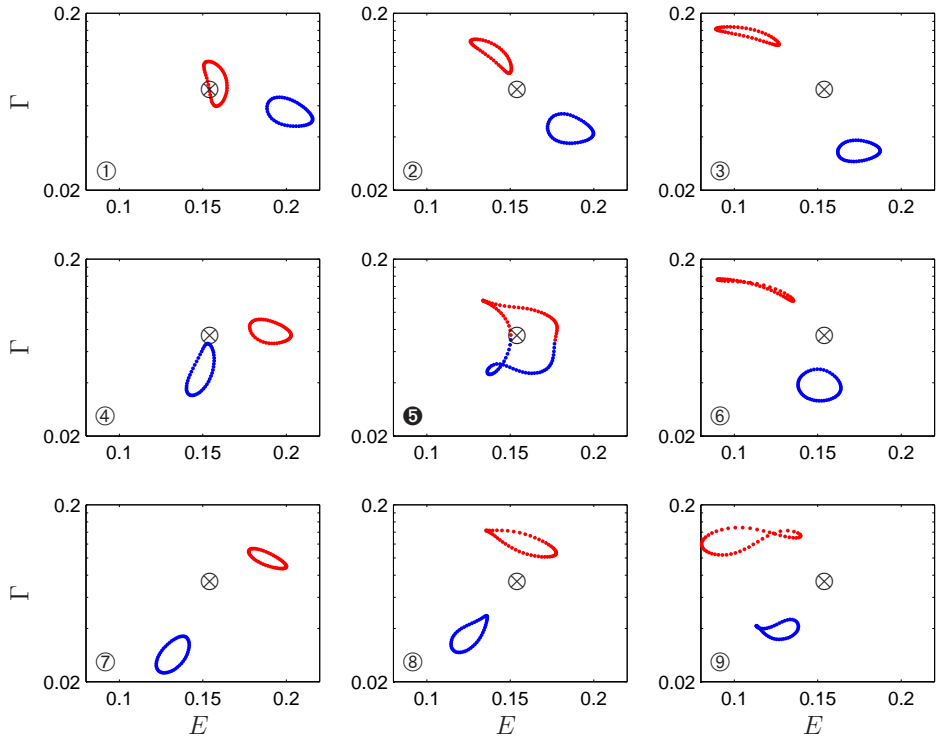}
 \caption{Eigenvalue trajectories in the complex energy plane $\mathcal{E}=E-\i \Gamma/2$ for a
          cyclic parameter variation of $F$ and $\phi$ for $\delta=1$, cf. also figure
          \ref{fig-mws1-EPcircle1-path}. The  energy of the EP is marked by $\varotimes$.}
 \label{fig-mws1-EPcircle1-E}
\end{figure}

Apart from paths that enclose the EP
$(1/F_{\mathrm{EP}},\delta_{\mathrm{EP}},\phi_{\mathrm{EP}})$ we are also interested in paths that do not enclose it, which can be controlled via the shifts $F_0$ and $\phi_0$.
Figure \ref{fig-mws1-EPcircle1-path} shows nine selected paths, the corresponding
eigenvalue trajectories in the complex energy plane are displayed in figure
\ref{fig-mws1-EPcircle1-E}.

We observe two qualitatively different kinds of behavior: 
Any closed path in parameter space that does not enclose the EP leads to eigenvalue trajectories in the complex energy plane that consist of two 
separated closed curves, one for the ground state and one for the first excited state.
Thus the eigenvalues always remain on the same Riemann sheet.
A different result is obtained for the path \ding{206} that encloses the EP. Instead of two separate eigenvalue 
curves in the complex energy plane we now find a single closed curve consisting of the complex eigenvalues of 
both eigenstates involved.
This property is typical of exceptional points \cite{Berr84b} and provides a useful criterion for proving the 
existence of an exceptional point in experiments (see e.~g.~\cite{Cart07,Cart08}). After two full cycles the 
initial eigenvalue is recovered, i.e.~the eigenvalues vary with a period of $4\pi$ in parameter space. 

It is also instructive to look at the eight outer subplots in figure \ref{fig-mws1-EPcircle1-E}
in the cyclic sequence
\begin{equation}
\text{\ding{192}}
\rightarrow \text{\ding{193}} \rightarrow \text{\ding{194}} \rightarrow \text{\ding{197}} 
\rightarrow \text{\ding{200}} \rightarrow \text{\ding{199}} \rightarrow \text{\ding{198}}
\rightarrow \text{\ding{195}} \rightarrow \text{\ding{192}}\,.
\end{equation}
Comparing the eigenvalue trajectories for path \ding{192} and path \ding{195} we observe 
that the assignment of the two trajectories to the ground state  (\textcolor{blau}{blue}) 
and the first excited state (\textcolor{rot}{red}) is interchanged. This can be interpreted as
as closed path in parameter space that also encloses the EP and hence moves from
one Riemann sheet to the other.

\begin{figure}[b]
\centering
            \includegraphics[height=5.0cm, angle=0]{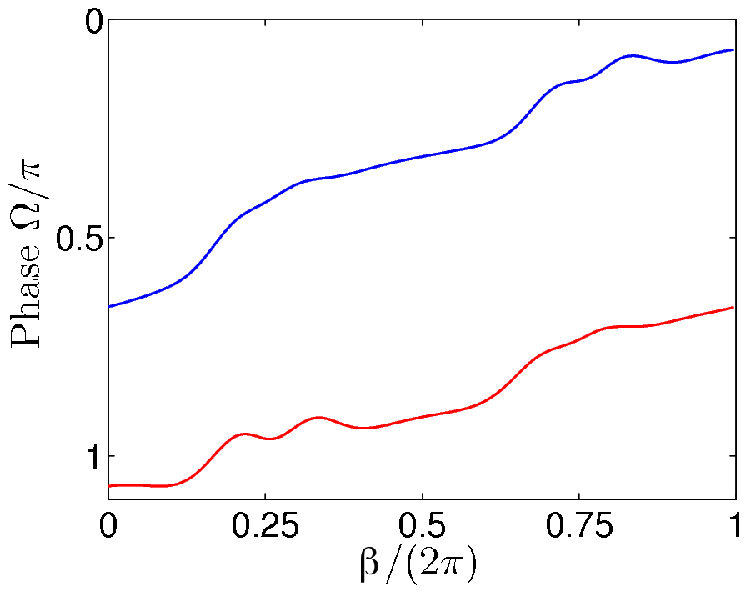}
	    \hspace*{4mm}
            \includegraphics[height=5.0cm, angle=0]{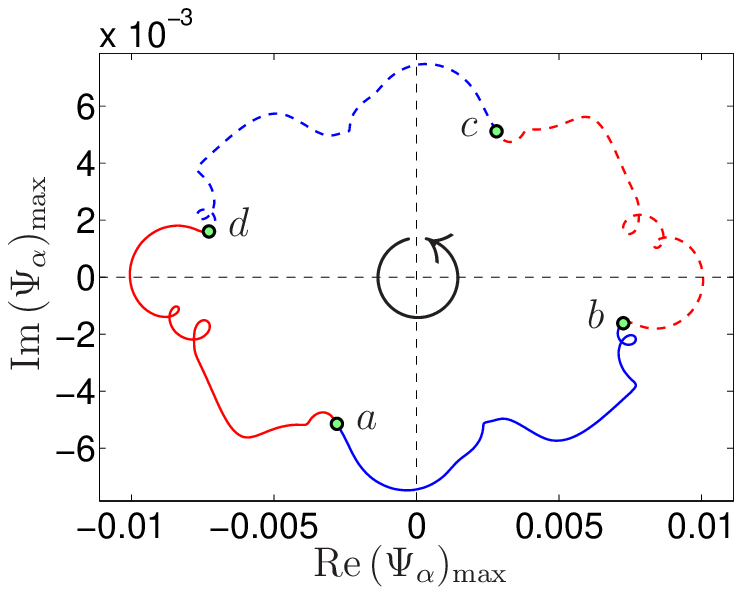}
\caption{Component $(\Psi_{\alpha})_{\mathrm{max}}= |(\Psi_{\alpha})_{\mathrm{max}}| 
\,\re^{-\ri 2 \pi \Omega }$ of the ground state \textcolor{blau}{(blue)} and the excited state
\textcolor{rot}{(red)} for a
cyclic parameter variation around the EP (\ref{math-mws1-ep}). Shown is the
phase $\Omega$ (left panel) as well as the real and imaginary parts (right panel).}
\label{fig-mws1-EPcircle1-ev}
\end{figure}

Apart from the eigenvalues we are also interested in the eigenfunctions. In order 
to monitor their behavior, it is sufficient to consider only one complex number, namely  
the projection of each wavefunction on one of the basis vectors in the plane-wave basis
chosen for the calculations. 
Here we choose the projection 
$(\Psi_\alpha)_{\mathrm{max}}$ with the maximum value of $|\Psi_{\alpha}|^2$;
the other projections show a qualitatively similar behavior.
In figure \ref{fig-mws1-EPcircle1-ev} we display the 
phase $\Omega$ (left panel) as well as the real and imaginary parts (right panel)
of 
$(\Psi_{\alpha})_{\mathrm{max}}= |(\Psi_{\alpha})_{\mathrm{max}}| \,\re^{-\ri 2 \pi \Omega }$
for the ground and first excited state for a
cyclic parameter variation around the EP, i.e.~when the phase $\beta$ in
parameter space varies form $0$ to $2\pi$. The figures
reveal the expected behavior:
After one complete cycle around the EP the phases $\Omega$ of the two components
are interchanged, however with 
a phase change of $\pi$, i.e.~a different sign. 
In the right panel, showing the eigenvector components in the complex plane, one observes the resulting 
$8\pi$-periodicity.   
Starting at the point {\rm (a)} (ground state) and {\rm (d)}  (first excited state) 
for $\pi=0$, the  selected components $(\Psi_{1})_i$ 
reaches points {\rm (b)} and {\rm (a)} after one cycle, where  {\rm (b)} corresponds to 
the sign-changed component $(\Psi_{2})_i$ at {\rm (d)}. The dashed lines show the 
development for three further cycles. 

\begin{figure}[t]
\centering
\includegraphics[height=5.5cm, angle=0]{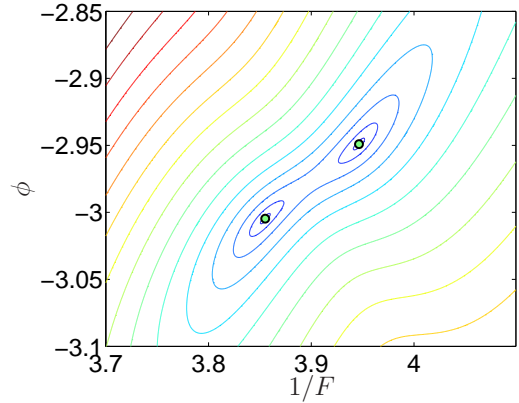}
\caption{Difference of the complex energies $|\mathcal{E}_2(1/F,\phi)-\mathcal{E}_1(1/F,\phi)|$ as a
contour plot for $\delta=2.3$. The marked minima correspond to the configuration of the EPs. The absolute values of the energy difference are color coded. Blue indicates for zero, red the maximum value. }
\label{fig-mws1-energydiff}
\end{figure}

\begin{figure}[b]
\centering
            \includegraphics[height=5.5cm, angle=0]{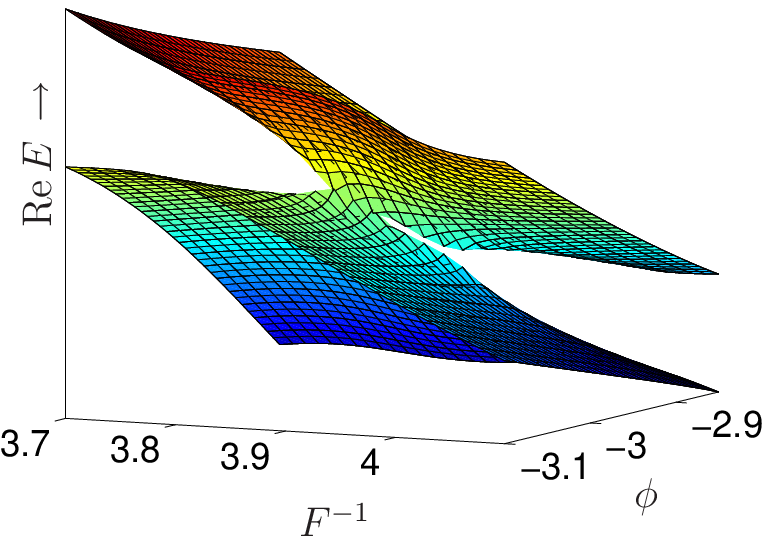}
            \includegraphics[height=5.5cm, angle=0]{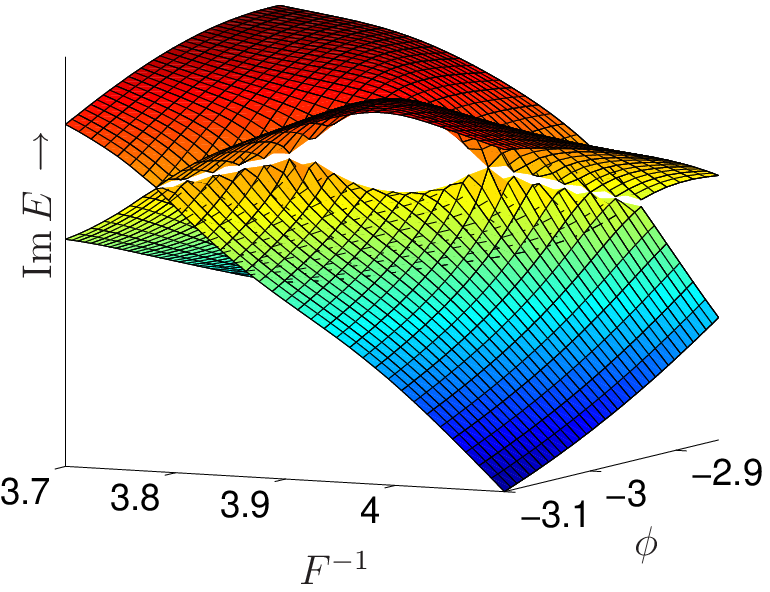}
\caption{Energy surfaces $\mathcal{E}_1,\,\mathcal{E}_2$ in the real- (left panel) and
imaginary part (right panel) for $\delta=2.3$.} 
\label{fig-mws1-energyplane}
\end{figure}

The periodicity of the eigenvalues and eigenfunctions can be summarized in the following diagrams 
(cf.~\cite{03crossing})
\begin{eqnarray}
 \underbrace{
\left\{ {\mathcal{E}_1 \atop \mathcal{E}_2} \right\}
\stackrel{2\pi}{\circlearrowleft} 
\left\{ {\mathcal{E}_2 \atop \mathcal{E}_1} \right\}
 \stackrel{2\pi}{\circlearrowleft} 
\left\{ {\mathcal{E}_1 \atop \mathcal{E}_2} \right\}
 }_{\displaystyle 4\pi},
\end{eqnarray}
\begin{eqnarray}
 \underbrace{
\left\{ {\Psi_1 \atop \Psi_2} \right\}
\stackrel{2\pi}{\circlearrowleft} 
\left\{ {-\Psi_2 \atop \Psi_1} \right\}
 \stackrel{2\pi}{\circlearrowleft} 
\left\{ {-\Psi_1 \atop -\Psi_2} \right\}
 \stackrel{2\pi}{\circlearrowleft} 
\left\{ {\Psi_2 \atop -\Psi_1} \right\}
 \stackrel{2\pi}{\circlearrowleft} 
\left\{ {\Psi_1 \atop \Psi_2} \right\}
 }_{\displaystyle 8\pi} \ ,
\end{eqnarray}
where $\Psi_1$, $\Psi_2$ are the Wannier-Stark eigenstates involved in the formation of the EP and 
$\mathcal{E}_1$, $\mathcal{E}_2$ the corresponding eigenenergies. 

Finally, looking again at figure \ref{fig-mws1-energyls}, we observe that in the vicinity of the
parameter values $1/F=4$, $\delta=2.3$ and $\phi=-3$ two exceptional points approach each other.
This region is magnified for $\delta= 2.3$ in figure \ref{fig-mws1-energydiff}, which shows the 
energy difference 
$\Delta \mathcal{E} = |\mathcal{E}_{1}-\mathcal{E}_{2}|$ in dependence on $1/F$
and $\phi$.
It is instructive to have a closer look at the
surfaces of the real and imaginary part of the eigenenergies 
in parameter space in this region. This is illustrated in
figure \ref{fig-mws1-energyplane}, which shows the Riemann surfaces of the complex energy in 
the vicinity of these two EPs. The observed local behavior resembles the 
``double-coffee-filter'' scenario typical of unfolding a diabolic point into two EPs 
\cite{03crossing,Berr03,Seyr05,Kiri05,Kiri09}. 
Note that the apparent gaps in the Riemann surfaces are an artifact of the finite discretization 
of the parameter space.

\begin{figure}[htb]
\centering
 \includegraphics[height=6cm, angle=0]{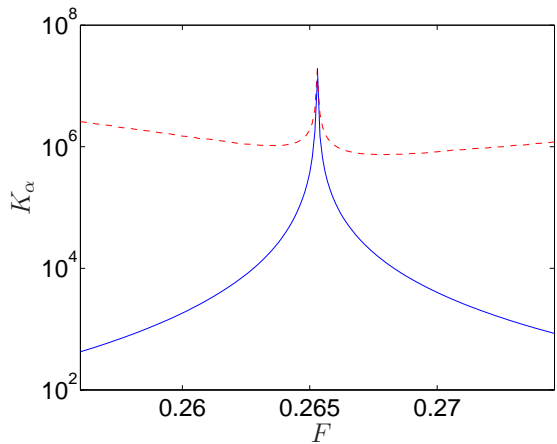}
 \caption{Petermann factor $K_{\alpha}$ of the ground state ($K_1$,
\textcolor{blau}{blue}) and the first excited state ($K_2$, \textcolor{rot}{red}) for a
double-periodic Wannier-Stark system. The parameters are in the vicinity of the EP
(\ref{math-mws1-ep}).}
 \label{fig-cnl-ep-1a}
\end{figure}

\subsection{The Petermann factor}
\label{sec-petermann}
For a non-hermitian system, one has to distinguish between right and left Wannier-Stark states,
i.e.~the eigenstates of  $H$ for eigenvalue $E_\alpha$ and of $H^\dagger$  for 
eigenvalue $E_\alpha^*$,
denoted as $|\Psi_{\alpha}\rangle$ and $|\widetilde{\Psi}_{\alpha}\rangle$, respectively. 
These states form a bi-orthogonal set, i.e.
$\braket{ \widetilde{\Psi}_{\alpha}}{ \Psi_{\alpha'}}=0 $ for 
$\alpha \ne \alpha`$\,. 
At an exceptional point, two eigenstates of $H$ coincide, as well as the corresponding
eigenstates of  $H^\dagger$, with the consequence that the scalar product
$\braket{ \widetilde{\Psi}_{\alpha}}{ \Psi_{\alpha}}$ vanishes.
A useful quantity in 
this context is the Petermann factor\cite{Pete79,Berr03a}
\begin{equation}
 K_{\alpha} \equiv
\frac{\braket{\widetilde{\Psi}_{\alpha}}{\widetilde{\Psi}_{\alpha}}\braket{\Psi_{\alpha}}{\Psi_{
\alpha}}}{\big|\braket{ \widetilde{\Psi}_{\alpha}}{ \Psi_{\alpha}}\big|^2}\,,
\label{mws1-petermann}
\end{equation}
which provides a measure of the overlap between left and 
right eigenvectors. Note that $K_{\alpha}$ is independent of the normalization of the states
and diverges at the EP. The Petermann factor provides a convenient tool for localizing
an exceptional point.

 Figure \ref{fig-cnl-ep-1a} shows
the Petermann factor of the ground state and the first excited state in dependence on the 
field strength $F$. The modulation $\delta$ and phase $\phi$ were chosen according to the 
configuration of the EP (\ref{math-mws1-ep}). The offset between the two curves is caused 
by the difference in the decay rates of the two respective states. At the position corresponding 
to the EP both Petermann factors $K_{1,2}$ show pronounced peaks.  Due to the discretization of the field 
strength and the fact that the configuration (\ref{math-mws1-ep}) is only very close to but 
not exactly in the EP, the numerical value of $K_\alpha$ remains, of course, finite.

\section{Nonlinear crossing scenarios}
\label{sec-nlcross}

One of the most promising realization of Wannier-Stark systems, both theoretically and experimentally, 
is based on atomic Bose-Einstein condensates in tilted optical lattices. Near the zero temperature limit 
these systems can be described by a nonlinear Schr\"odinger equation (NLSE) or  Gross-Pitaevskii 
equation (GPE) (see e.~g.~\cite{Pita03,Park98,Dalf99,Legg01})
\begin{equation}
 \left[ -\frac{\hbar^2}{2m} \frac{\rd^2}{\rd x^2} + V(x) + Fx + g|\Psi(x)|^2 \right] \Psi(x) =
\mu \Psi(x)
\label{math-num-gpe-ntd}
\end{equation}
in a mean-field approach, where the nonlinear term $g|\Psi(x)|^2$ takes into account the interaction 
between the condensate particles. The interaction between the particles can either be repulsive 
($g>0$) or attractive ($g<0$). In the case of Wannier-Stark resonances, the chemical potential $\mu=M-\ri 
\Gamma/2$ is complex where the real part $M$ describes the position of the resonance and the imaginary part 
$\Gamma/2$ accounts for the decay rate. 

\begin{figure}[t]
\centering
\includegraphics[height=5.0cm, angle=0]{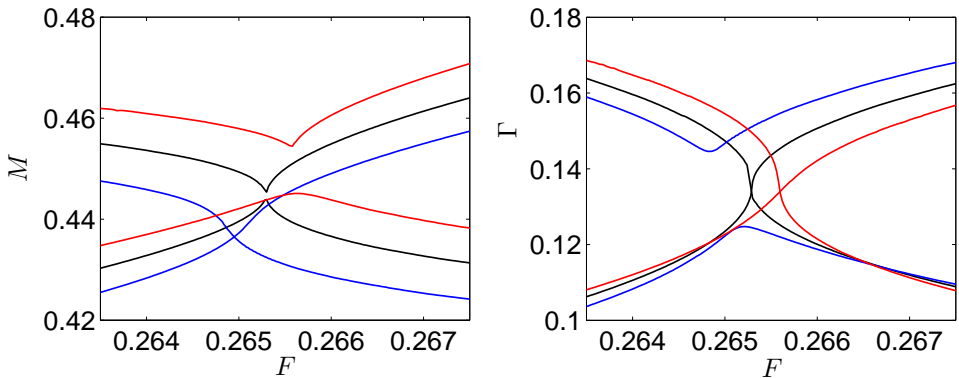}
\caption{Crossing behavior of the chemical potential $\mu=M-\ri \Gamma/2$ 
of the two most stable resonances
for  $g =
\textcolor{blau}{-0.02 \text{ (blue)}},\, 0\text{ (black)}$ and $\textcolor{rot}{0.02\text{ (red)}}$
in the vicinity of the  EP (\ref{math-mws1-ep}) for $g=0$, where
real parts (left) and imaginary parts (right) are degenerate at $F=0.2653$.}
\label{fig-cnl-ep-realimag}
\end{figure}

Up to now, complex eigenvalue crossing scenarios and exceptional points in the context of nonlinear 
systems have mostly been investigated by means of simple nonlinear non-hermitian two-level models 
\cite{06nlnh,06zener_bec,08PT}, three-level systems  \cite{06level3}, 
or to some extent analytically solvable model systems like one-dimensional $\delta$- and
$\delta$-shell potentials \cite{04nls_delta} or
multiple-barriers \cite{09ddshell} of a delta-comb \cite{09dcomb}. An exception is
the discovery of EPs for atomic gases with an attractive  $1/r$-interaction 
reported in \cite{Cart08}.

The numerical calculation of nonlinear Wannier-Stark resonances is a nontrivial task
because most methods established for the linear system cannot be
straightforwardly adapted since the superposition principle is no longer valid.
In \cite{Wimb06,07nlres} nonlinear Wannier-Stark resonances of the ground ladder
in single-periodic and double-periodic systems were calculated using a complex
scaling procedure developed in \cite{Mois05,Schl06a}. Here we use a method based
on complex absorbing potentials in position space \cite{10nlret} as it is the only procedure so far that 
has successfully been used to calculate Wannier-Stark resonances of excited ladders.
In the following we want to briefly discuss the impact of the nonlinear interaction on eigenvalue 
crossing scenarios and exceptional points for the bichromatic WS-system discussed
in the preceding sections.

As a first example, figure \ref{fig-cnl-ep-realimag} shows
the real and imaginary parts of the eigenvalues of the chemical potential $\mu$ in dependence on 
the static field strength $F$ for a WS-system again in the vicinity of the EP (\ref{math-mws1-ep}).
The black curves are the real and imaginary parts of the resonance energies
of the two most stable states for the non-interacting system ($g=0$) which cross
at $F=0.2653$. A moderate nonlinearity removes this degeneracy 
and we observe  a type I crossing (cf.~section \ref{s-Intro}) for a repulsive interaction 
($g=+0.02$, red curves) and a type II crossing for an attractive interaction 
($g=-0.02$, blue curves) in agreement with the results already
briefly reported in \cite{10nlret}.

\begin{figure}[htb]
\centering
 \includegraphics[height=6cm, angle=0]{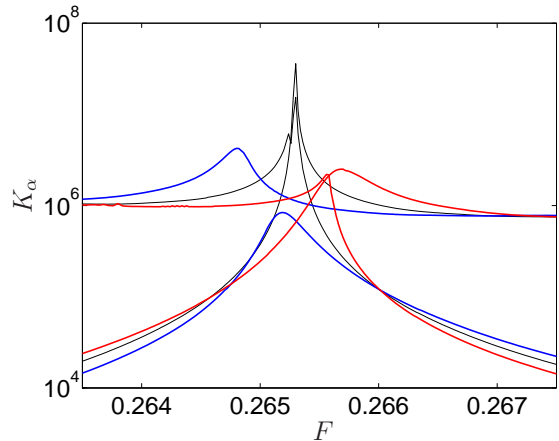}
 \caption{Petermann factor of the ground state $K_1$ and of the first excited state $K_2$ for a
parameter variation of $F$ around the EP (\ref{math-mws1-ep}) of the linear system. The interaction strength was 
chosen as $g = \textcolor{blau}{-0.02 \text{ (blue)}}, 0 \text{ (black)}$ and $\textcolor{rot}{0.02 \text{ (red)}}$.}
 \label{fig-cnl-ep-petermann}
\end{figure}

This indicates that the position of the EP is shifted by the 
nonlinearity in opposite directions for positive or negative values of 
$g$, assuming that the existence of the 
EP is not affected by weak interactions.
Figure \ref{fig-cnl-ep-petermann} displays the Petermann factors $K_1$ and
$K_2$ for the resonance states shown in figure \ref{fig-cnl-ep-realimag}.
The Petermann peaks for $g=\pm 0.02$ are now strongly reduced and shifted to stronger
fields for repulsive and weaker fields for attractive interactions. Moreover,
the two maxima are also shifted
relative to each other: The maxima of the excited state are more affected
by the nonlinearity than the ground state. 

\begin{figure}[b]
\centering
 \includegraphics[height=6cm, angle=0]{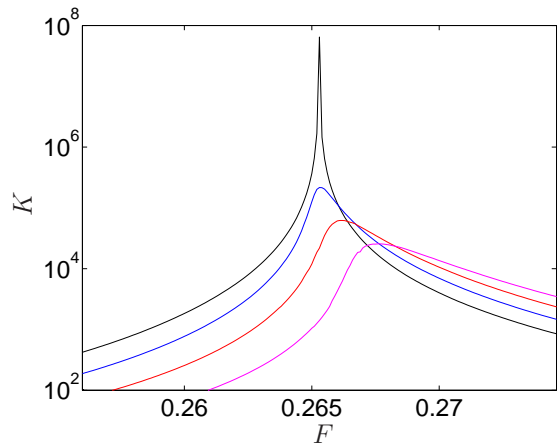}
 \caption{Petermann factor $K_1$ of the ground state for a variation of $F$ around the configuration of the EP
(\ref{math-mws1-ep}) in the linear case. The interaction strength is chosen as $g = 0, 
\textcolor{blau}{-0.1 \text{ (blue)}}, \textcolor{rot}{-0.2 \text{ (red)}}$ and
$\textcolor{magenta}{-0.3 \text{ (magenta)}}$.}
 \label{fig-cnl-ep-1b}
\end{figure}

Finally we have a closer look at the range of moderately increased attractive interaction 
strengths $g$ with $-0.3 < g < 0$. The corresponding results are shown in figure 
\ref{fig-cnl-ep-1b} for the ground state. With increasing $|g|$
we observe that the Petermann peaks get more and more ``blurred'' and become 
strongly asymmetric, indicating that the system moves away from the true 
degeneracy in parameter space.

An extension of the systematic search for EPs as described in section \ref{subsec-systematic} to nonlinear systems 
deserves clearly further studies.
In particular, it has to be checked if the exceptional point 
exists for all values of $g$, or if it vanishes when the nonlinearity 
exceeds a critical value. However, such studies are computationally 
demanding, in particular because of the convergence properties of the method for calculating 
nonlinear Wannier-Stark resonances. Thus substantial modifications of the presented numerical procedures 
are required. In this context the criterion of a vanishing inverse Petermann factor $1/K_{\alpha}$ for the 
configuration of an EP might help. One advantage of this criterion is that it only requires the eigenstates 
of a single ladder.

\section{Conclusion}
\label{sec-WS_conclusion}

This paper investigated the spectrum of an experimentally realizable model system for
decay in open quantum systems, namely a quantum particle in a tilted bichromatic lattice.
The analysis concentrated on the exceptional points (EPs) of the system, i.e. points in parameter 
space where both the complex eigenvalues and the eigenfunctions of two different eigenmodes coincide, 
which had so far mostly been considered for very simple toy models. An efficient method for finding EPs was 
presented and their location within the parameter space was discussed in detail as well as the properties of the corresponding degenerate eigenfunctions. 
It was demonstrated that in the limit case of a monochromatic lattice there can be no exceptional points for finite values of the tilt, because 
degeneracies of the eigenenergies can only occur for eigenstates of different ladders which cannot coincide as they localize in
different lattice sites. Furthermore the geometric phases (Berry phases) and the eigenvalue trajectories in the complex plane for closed paths in parameter
space were considered. The eigenvalue trajectories show the familiar behavior; they form a single closed curve for paths in parameter
space that enclose an EP but two distinct closed curves otherwise. In particular, the two crossing states are interchanged after 
a full cycle in parameter space enclosing an EP. 
Finally the case of an interacting Bose-Einstein condensate in a tilted bichromatic optical lattice was discussed within a mean-field approach
by solving the corresponding nonlinear Schr\"odinger equation or Gross-Pitaevskii equation. It
was found that the type of crossing, i.~e. if the real parts of the eigenenergy anti-cross while
the imaginary parts (decay rates) cross (Type I) or vice versa (Type II), depends on the sign of the interaction between the particles 
(i.e. whether it is attractive or repulsive). These crossing scenarios, as well as an analysis of the 
Petermann factor, which measures the overlap between left and right eigenstates and shows a pronounced peak 
in the vicinity of an EP, indicates that the EPs are shifted by the nonlinear interaction in a direction opposite to its sign.

\ack
We would like to thank Sandro Wimberger for fruitful discussions. Financial support by 
the Deutsche Forschungsgemeinschaft (DFG) via the Graduiertenkolleg 792 "Nichtlineare Optik 
und Ultrakurzzeitphysik" is gratefully acknowledged.
K.R. acknowledges support from a
scholarship of the Université Libre de Bruxelles.

\section*{References}
\bibliographystyle{/home/agkorsch/tex/bibtex/bst/unsrtot}
\bibliography{/home/agkorsch/tex/bibtex/bib/abbrev,/home/agkorsch/tex/bibtex/bib/publko,/home/agkorsch/tex/bibtex/bib/paper60,/home/agkorsch/tex/bibtex/bib/paper70,/home/agkorsch/tex/bibtex/bib/paper80,/home/agkorsch/tex/bibtex/bib/paper90,/home/agkorsch/tex/bibtex/bib/paper00,/home/agkorsch/tex/bibtex/bib/rest,/amd/home/aleph/elsen/DA/rev01/common}
\end{document}